# G$^3$: GENESIS SOFTWARE ENVIRONMENT UPDATE

*Nicolas Castagne*

ACROE
INPG, Grenoble, France
castagne@imag.fr

*Claude Cadoz*

ACROE & ICA laboratory
INPG, Grenoble, France
castagne@imag.fr

*Ali Allaoui*

ICA laboratory
INPG, Grenoble, France
castagne@imag.fr

*Olivier Tache*

ACROE
INPG, Grenoble, France
tache@imag.fr

**ABSTRACT**

GENESIS$^3$ is the new version of the GENESIS software environment for musical creation by means of mass-interaction physics network modelling. It was re-designed from scratch in hindsight of more than 10 years working on and using the previous version. We take the opportunity of this release to provide in this article an analysis of the specificities of GENESIS and an update on the features of the new version.

## 1. INTRODUCTION

In the past 15 years, physical modelling and simulation have been a major issue in Computer Music research, and have been used in some musical pieces [5]. Though, physical modelling is still mostly considered as a specialist's activity and musicians rarely practice themselves modelling. Physical models are most often available as "black boxes" (*e.g.* synthesis plug-ins) dedicated to certain categories of acoustic structures, and provided within non-physical-modelling software environments [10]. Indeed, only a few software environments, among which [3, 7, 8], have aimed at empowering musicians and composers with physical modelling (and not only with physical *models*), and the adoption of physical modelling by end users as a musical activity is still an issue. We assume this requires considering that physical modelling is a fundamentally new paradigm in Computer Music, that calls for the development of new creation processes and know-how [9], and the design of appropriate software means, with their associated dedicated interfaces.

At ACROE-ICA, we have carried out since the 90's a research dealing with both technological issues and musical aspects on GENESIS [3], a musician-oriented software environment for musical creation with physical modelling based on mass-interaction physics networks, as defined in the CORDIS-ANIMA [1] formalism.

As proved by the 50000 models [9] built within the GENESIS user group, physical modelling is in GENESIS not only a mean for sound synthesis, but becomes more generally a mean for musical creation, including the compositional activities. In particular, the "*compose (with) physical modelling*" process [2], that may involve tens thousands modules, demonstrates that physical modelling allows melding both sound synthesis and computer-aided composition, and that it can be relevant for dealing with compositional ideas.

The first version of GENESIS, GENESIS 1 (G1) was beta-released in 1995, and published in the early 2000's [3]. In hindsight of more than 10 years, we started in 2006 the building of a new version. GENESIS$^3$ (or G$^3$, pronounce "cube") was born in the early 2009.

The design process of G$^3$ aimed first at providing a more ambitious version, able to run transparently on Mac, Linux and Windows, and able to support the most demanding usages and creation processes. Another long term ambition was generalizing the involved software components to support, in the future, multi-sensory models and creations. A team in the research group has developed the Computer Graphics MIMESIS software for image and movement synthesis with physical modelling [6], and another team works on multisensory, hard real time physically based simulation featuring force-feedback devices. These teams got involved in the design process toward G$^3$, and their needs, existing software, and uses have been considered. Hence, G$^3$ is released also as the first element of an upcoming generic software suite for artistic creation with physical networks.

Taking the opportunity of G$^3$'s birth, this article provides an update on the GENESIS software environment. Many of the fundamentals introduced in G1 were confirmed, and are not discussed much in the following – see [3]. The article focuses more specifically on GENESIS core principles, and on the new features.

## 2. CORDIS-ANIMA IN GENESIS

GENESIS is based on the CORDIS-ANIMA physical modelling and simulation formalism [1]. So as G1, $G^3$ features the "uni-dimensional", or "topological" version of CORDIS-ANIMA: each physical variable (position, force) is computed along a single movement axis. A CORDIS-ANIMA physics network is a network made of elementary material modules `<MAT>` connected with physical interactions `<LIA>`. Each module stands for an elementary physical behavioural law: inertia, stiffness, viscosity, buffer interaction, non linear interaction, etc.

$G^3$ provides the same module as the former version – see [3]. 12 simple elementary modules are available, and 7 physical parameters and 2 initial state properties exist in GENESIS. No matter how complex are the designed models and how ambitious are the musical goals pursued, the user's work always relies on assembling instances of these 12 basic and intuitive physical modules in a network, setting their parameters, then simulating the model.

A major role of GENESIS, reified in many of the available features, is then to provide means to handle (generate, parameterize, organize, etc…) the designed physics network in order to meet some musical needs. But the user always can access (modify, parameterize, remove…) individually each elementary module in the model. This is necessary for the network topology (large regular networks are a rare case in GENESIS) as well as for the modules' parameters (homogeneous parts in the network can happen, but this is not a general rule).

The highly modular scheme based on a very small number of physical-only modules is one of the major signature of GENESIS. It differs noticeably from other computer music software paradigms. For example, letting the user manipulate a single waveguide element (a delay filter) would not make much sense. Also, in Modalys [7], an object is yet a complex component corresponding to a collection of modes, and "controllers" and "connections" are core elements that are non-physical. And, with the Tao physical modelling language [8], and though the simulation system is based on mass-spring elements, the user handles pre-constructed objects such as strings and membranes. Finally, one can note that within signal-based patching environments, the number of module types is higher.

As a vis-à-vis of the simplicity of the module types, the number of elementary modules in a complex GENESIS model can be very large. The new version may support hundred(s) thousands modules. This is another core signature of GENESIS. Models in other modular environments usually feature much less elements, though taken from a larger set of categories with more complex algorithms. The possibility of huge sets of independent modules in models has deep consequences throughout GENESIS. It impacts deeply the data structures, the required efficiency and the needed features.

## 3. DIRECT MANIPULATION AND ZOOMING

$G^3$ is centred on a graphical representation of the designed model on a 2D workbench, which features direct manipulation (Figure 1). The single movement axis is perpendicular to the workbench. The two dimensions of the workbench hence have no impact on the physical behaviour of the model during simulation. They are left for the user to organize freely the model. But position matters: users always carefully choose the placement of modules.

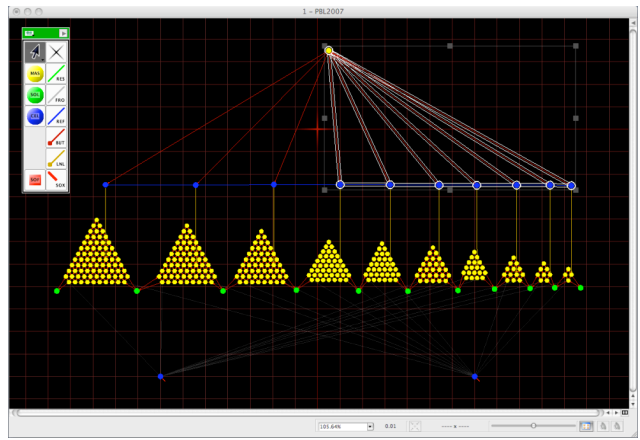

**Figure 1.** a workbench (here on the Mac)

$G^3$ improves the representation and the navigation features introduced in G1, by taking inspiration from the research in the field of *zoomable user interfaces* (ZUI) [11]. One can note that zooming is rarely a core feature in signal-oriented environments – where encapsulation is usually preferred, see next section. Conversely, zooming and navigation are core means in GENESIS.

In $G^3$, an heuristic algorithm adapts the representation of modules and models based on the current viewpoint (zoom factor, position in the whole model, proportion of visible modules, etc.). Direct manipulation features, such as pointing a module, selecting modules, etc, are, on their turn, adapted to the representation.

To let the user annotate their models, $G^3$ allows placing *bench notes* within models. A bench note is a textual item encoded in HTML, including hyperlinks. Thanks to a dedicated series of URL schemes, such links can activate various features throughout the interface, and refer to any item in the model, allowing for example to quickly select the referred items.

## 4. LABELS AND SUB-NETWORKS

The need of huge sets of modules calls for dedicated features able to handle subset of modules and to reduce the user's cognitive load. Facing such needs, features that are usually implemented are *grouping* (i.e. : grouping a set of elements to let the user handle it as a single super-element) and *encapsulation* (i.e.: hide the inner complexity of a group by providing an iconic representation for it). However, grouping and encapsulation are not appropriate for GENESIS.

First, physics mass-interaction networks makes it difficult to strictly isolate a sub-network. By nature any parts interacts bi-directionally with other parts. While designing, each part regularly requires to be modified, either in its network topology or in the parameters of some modules, in order to be adapted to other interacting parts. Hence, fundamentally, mass-interaction networks modelling do not match a strict tree-like structural approach when organizing a complex model. A model should always be seen as a "whole".

Second, the subsets of modules a user needs to consider most often overlap with each others. For example, at one time he/she could need to handle modules that corresponds to a "sub-object" (e.g.: a "string"); but during another phase of the modelling process, he/she could need to consider some modules in this sub-object along with other modules in another part of the model (e.g.: parameterize globally this new subset that spans over various sub-objects sharing some properties).

Given these observations, instead of grouping and encapsulation, $G^3$ builds upon the notion of "module sets" in G1 [3], and upon the concept of label in MIMESIS [6] to introduce the *labelling system*.

In $G^3$, we call a *label* a string that targets a module. Each module has a unique permanent label given by the system upon creation, but any number of user-defined label(s). User-defined labels can include "/", as separators. Separators are taken into account by the labelling system to regroup labels on the basis of their syntactic proximity. We then call a *sub-network* any set of modules that share labels with the same radical.

For example, if 3 modules are labeled using `/myString/extremities/1`, `/myString/extremities/2` and `/myString/aModule`, then `/myString` refers to a sub-network made of 3 modules, and `/myString/extremities` refers to another sub-network made of two modules.

The user can define as many labels for a module as there are contexts, or edition tasks, in which the module may be involved. Symmetrically, a module may belong to as many sub-networks as needed. Put differently, instead of structuring the modules in a model, the labelling system allows organizing modules' names in an oriented graph, in which nodes target sub-networks, and leaves target modules. The labelling system, with its interface, hence offers a particularly flexible, but not structurally oriented, mean to let the user deal with overlapping sets of modules.

## 5. THE PNSL LANGUAGE

The experience of our research group with mass-interaction networks lead us to consider that both *direct manipulation* and *textual interaction* (programming) have fundamental interests in the context of modelling. As an endpoint of the series of languages implemented in the laboratory, in particular by the MIMESIS team, $G^3$ introduces the user-oriented *Physics Network Scripting Language*. An article on PNSL is in preparation, and the following only provides a short introduction.

PNSL is a *modelling language*, aimed at supporting the modelling activity with mass-interaction physics networks, as defined in CORDIS-ANIMA. PNSL is also a *scripting language*. As such, conversely to other physics-based-language such as PML [4], but more like Modalys or Tao [8, 9], PNSL was not design to encode the state of a given model, but to let the user "program" the modelling activity. PNSL is noticeably *generic* in the context of physics network: any physics network may be designed with PNSL, no matter what its dimensionality is (1D, 2D, 3D), and the categories of phenomena to be generated are (sound, movement, animated images…). Hence, PNSL will be a major element in the upcoming software suite for physical modelling for the Arts (see introduction). Finally, PNSL has enough expression power and efficiency to support very large models, with hundred thousands modules. In particular, any PNSL command allows handling any module individually, but also numerous modules globally, thanks to *Label Picker Expressions*, and their pattern matching mechanism.

Technically speaking, PNSL is built upon Tcl. It groups 67 commands spanned over 13 packages dedicated to the handling of a category of data: module creation, label handling, physical parameter manipulation, etc.

In the first release of $G^3$, in addition to an extensive internal use of PNSL in the core mechanisms, the language is provided to the user as a macro language. Executing a script (or *macro*) modifies the state of the model at hand, which is immediately visible in the interface. In the script window, the user can select over pre-defined scripts in the included *PNSL library*, or create, edit and execute its own scripts.

## 6. SIMULATION

In GENESIS, simulation is not only a mean to obtain the final synthesized sound, but more importantly a major mean in the modelling activity itself. Users constantly switch between modelling and simulation in order to evaluate his/her modelling acts. In $G^3$, the modelling workspace (the workbench) is separated from the simulation workspace (the simulation window). User activities in both space deeply differ, and are of comparable importance.

We deeply redesigned the embedded off-time simulation engine in a highly multithreaded architecture, and its associated simulation window, so as to provide a configurable workspace, depending on the category of phenomena the user is interested in. The $G^3$ simulation window hence provides various layouts, allowing to display a 2D+1 graphical visualization of the model while simulated, the waveform of the generated sound file, or all of them, etc. Figure 2 provides a snapshot of the simulation window with its full layout. There, the user can seamlessly listen to the sound, launch the 2D+1 visualization process, pursue the simulation, etc.

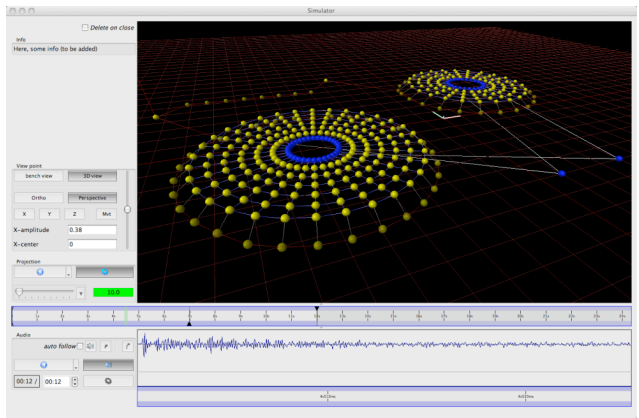

**Figure 2.** Full layout of the simulation window.

Finally, $G^3$'s simulator was designed to allow using various simulation engines. This will allow switching between the embedded off-time CORDIS-ANIMA simulator and the on line, hard real time, interactive and multisensory simulator, featuring force feedback devices, developed in the research group.

## 7. CONCLUSION

The core of the GENESIS software environment was confirmed and refined through more than 10 years of evolution and usage of G1. It corresponds with new music creation processes, in which physical modelling is central. With $G^3$, we hope we achieved both a stabilization of GENESIS' principles, and a more usable and powerful version.

The birth of $G^3$, on its turn, opens for many years a new series of evolutions. Future research direction regarding the software aim in particular at completing the high level edition tools it features, and better supporting the most innovative uses. Recent software experiments have been conducted, especially featuring the PNSL language, to allow extending the workbench usages by assigning some musically significant meaning to the workbench 2 dimensions. Evolved parameter edition, based on the possibility of defining mathematical relations amongst sets of parameters to meet various goals, is another work direction. Finally, a major research axis, already started in the laboratory, aims at enhancing the graphic-textual (or "workbench-PNSL") collaboration in GENESIS, toward a deeper bi-modal software melting appropriately direct manipulation and language-based modelling.

## 8. ACKNOWLEDGEMENTS

We thank our colleagues A Luciani, JL Florens, and M Evrard for their deep implication in the design process, and the beta-testers for their feedbacks. Many thanks are also due to the developers of the open source software used in $G^3$: Tcl, libsndfile, portaudio.